# ADAPTING MAC 802.11 FOR PERFORMANCE OPTIMIZATION OF MANET USING CROSS LAYER INTERACTION

Gaurav Bhatia[1] and Vivek Kumar[2]

gbhatia13@gmail.com[1], vivekdcg@gkvharidwar.org[2]

Research Scholar[1], Supervisor[2]
Department of Computer Science,
Gurukul Kangri Vishwavidyalya,
Haridwar-249404, India.

## ABSTRACT

*In this research, we study the optimization challenges of MANET and cross-layer technique to improve its performance. We propose an adaptive retransmission limits algorithm for IEEE 802.11 MAC to reduce the false link failures and predict the node mobility. We implemented cross layer interaction between physical and MAC layers. The MAC layer utilizes the physical layer information for differentiating false link failure from true link failure. The MAC layer adaptively selects a retransmission limit (short and long) based on the neighbour signal strength and sender node speed information from the physical layer. The proposed approach tracks the signal strength of each node in network and, while transmitting to a neighbour node, if it's received signal strength is high and is received recently then Adaptive MAC persists in its retransmission attempts. As there is high probability that neighbour node is still in transmission range and may be not responding due to some problems other then mobility. In this paper, we evaluate the performance of MANET and show that how our Adaptive MAC greatly improves it. The simulation is done using Network Simulator NS-2.*

## 1. INTRODUCTION

Mobile Ad hoc Networks (MANET) can be defined as autonomous systems of mobile nodes connected via wireless links without using an existing network infrastructure or centralized administration [1]. The nodes composing a MANET are free to move and to organize themselves arbitrarily and thus the topology of the network may change rapidly and unpredictably. In multihop ad hoc networks, every node acts also as a router and forwards each others' packets to enable the communication between nodes not directly connected by wireless links. The benefits and commercial potentials of the ad hoc architecture have attracted considerable attention in different application domains. Unfortunately, the layered open system architecture (OSI) does not seem to support these requirements. The layered architecture is remarkably successful for networks made up of wired links, where the key assumptions and abstraction boundaries work well. The strict layering approach reveals to be suboptimal in many application domains of MANET [2]. The main drawback of the ISO/OSI model is the lack of cooperation among layers: each layer works in isolation with little information about the network. Moreover, the strict modularity does not allow designing joint solutions optimized to maximize the overall network performance.

Cross layer interaction is an emerging proposal to support flexible layer approaches in MANET. Generally speaking, cross layer interaction refers to protocol design done by allowing layers to exchange state information in order to obtain performance gains. Protocols use the state information flowing throughout the stack to adapt their behavior accordingly. The cross layer interaction introduces the advantages of explicit layer dependencies in the protocol stack, to cope with poor performance of wireless links and mobile terminals, high error rates, power saving requirements, quality of services etc. Many interesting cross layer design solutions have been proposed in literature [3, 4, 5], together with some critical works addressing the risks of an unbridled cross layer design leading to uncoordinated interactions, fluctuations, and system instability.

IEEE 802.11 MAC protocol is the standard for wireless networks. It is widely used in simulations in the research for mobile ad hoc networks. It reports a link failure if it is not able to communicate with another node in fixed retransmission attempts. If the destination node is in transmission range and was not responding due to reason other than node mobility then this link failure is not true. Our idea is to use cross layer interaction for adapting the retransmission attempts to reduce the link failures.

## 2. OVERVIEW OF IEEE 802.11 MAC [6]

IEEE 802.11 MAC provides the basic access method based on the CSMA/CA (Carrier Sense Multiple Access with Collision Avoidance) scheme to access the channel. According to this scheme, when a node receives a packet to be transmitted, it first listens to the channel to ensure no other node is transmitting. In order to detect the status of the medium, IEEE 802.11 performs carrier sensing at both the physical layer, referred to as the physical carrier sensing and at the MAC layer, referred to as the virtual carrier sensing.

A virtual carrier sensing mechanism (Figure 1) is done via the use of two control packets (RTS/CTS) as follows: A source node ready to transmit senses a medium, if the medium is busy then it defers. If the medium is free for a specified time called *Distributed Inter Frame Space* (DIFS) then it sends a *Request to Send* (RTS) packet towards the destination. All other nodes that hear the RTS then update their *Network Allocation Vector* (NAV), which indicates the amount of time that must elapse until the current transmission session is complete and the channel can be sampled again for idle status. The destination node, upon reception of the RTS responds with another short control packet *Clear to Send* (CTS). All other nodes that hear the CTS packet also defer from accessing the channel for the duration of the current transmission. This means that, the channel is marked busy if either the physical or virtual carrier sensing mechanisms indicate the channel is busy. The reception of the CTS packet at the transmitting node acknowledges that the RTS/CTS dialogue has been successful and the node starts the transmission of the actual data packet after a specified time, called the *Short Inter Frame Space* (SIFS) and then transmits the packet. Otherwise, it chooses a random back-off value and retries later.

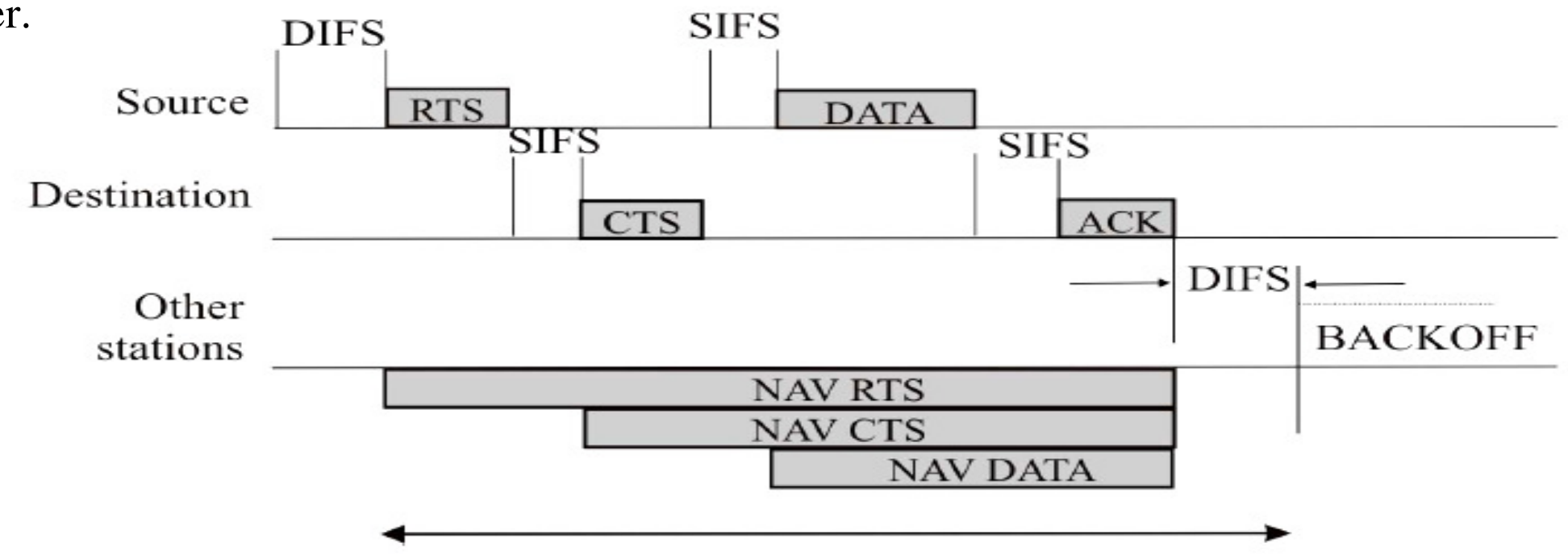

Figure1. Virtual Carrier Sense and Data Transmission in IEEE 802.11 MAC

If the receiver gets the packet without error, it initiates the transmission of Acknowledgement packet after a SIFS towards the sender. The SIFS is shorter than DIFS in order to give priority to the receiving station to over other possible stations waiting for transmission. However if the Acknowledgement packet is not received at the sender within a certain time, the Data packet is presumed to have been lost and a retransmission is scheduled. Similarly, if CTS control packet is not received when the MAC sent the RTS then MAC resends the RTS up to predefined limit. The RTS/CTS mechanism has two retry limits associated with it: Short Retry Limit (SRL) and Long Retry Limit (LRL).

The SRL is associated with the transmitting node sending out the RTS control packet and receiving back a CTS control packet from the destination. When the transmitting node sends out an RTS to the destination, and, after a small duration of time, if the transmitting node does not receive back CTS from the destination, the transmitting node begins retrying to send the RTS. This retransmission is done up to a certain number of attempts known as the SRL which has a default value statically set at 7. The amount of resend attempts is recorded in the *ssrc* counter variable within each packet. These variables are compared to SRL to decide whether to retry sending the packet or to discard the packet. Once the SRL is reached, the corresponding packet is discarded, and the transmitting node believes the destination node to no longer be accessible.

The LRL is associated with the transmission of a Data packet and an Acknowledgement packet. After the source sends its data packet, it then awaits a small duration of time to receive back an Acknowledgement packet. If the duration of time expires without receiving the ACK packet back, the transmitting node will then attempt to resend the data packet. This retransmission is done up to a certain number of attempts known as the LRL which has a default value statically set at 4. The amount of resend attempts is recorded in the each packet's *slrc* counter variable. This variable, similar to the SRL counter variable, is compared to the LRL to decide to either continue trying to resend the data packet or to discard it and move on to the next packet in the queue. If the packet cannot be sent successfully to the next hop within these limits (SRL, LRL), it would be dropped and trigger a link failure.

## 3. BACKGROUND AND MOTIVATION

We performed the comprehensive study of link failures in MANET and found that the main reasons for link failures are mobility, interference and congestion.

**3.1 Mobility.** In MANET, each node is free to move while communicating with other nodes. As one mobile node moves out of the other's transmission range, mobility in ad hoc networks causes frequent link failures, which in turn results in route failure and packet losses [1].

**3.2 Congestion.** Congestion (overload) may give rise to buffer overflow and increased link contention, which degrades MANET performance. As a matter of fact, [7] showed the capacity of wireless ad hoc networks decreases as traffic or competing nodes arise. If the network is heavily loaded, it is more likely that congestion dominates packet losses which in turn results in link failure.

**3.3 Interference.** The interference can be due to collision (by hidden node, extended hidden node and exerted hidden node), exposed node problem (self interference), TCP aggressive window mechanism.

**3.3.1 Hidden Node.** A hidden node is one which is within the interfering range of the intended destination but out of the sensing range of the sender. Hidden nodes can cause collisions on data transmission.Two nodes, out of each others' radio range, simultaneously try to transmit data to an intermediate node, which is in radio range of both the sending nodes. None of the sending nodes will be aware of the other node's transmission, causing a collision to occur at the intermediate node [1]. The hidden node interference can be avoided by using the RTS-CTS handshake method of 802.11 MAC.

**3.3.2 Extended Hidden Node.** According to the IEEE 802.11, the interfering range is more than two times the size of the sensing range. An extended hidden node is one which is out of sensing range of the sender but within the interfering range of sender [8]. Thus the RTS-CTS handshake method (virtual carrier sensing) cannot prevent all interference.

**3.3.3 Exerted Hidden Node.** IEEE 802.11 MAC protocol offers a lower throughput in the presence of heterogeneous network. In ad hoc wireless networks, we deal with nodes that have different power capabilities, hence, there is a considerable likelihood to transmit with different power levels. The principal cause of this interference is that high power nodes cannot sense the RTS/CTS dialog among low power nodes [9]. Thereby, the hidden terminal problem is exacerbated, which provokes more link failures.

**3.3.4 Exposed Nodes.** An exposed node problem is caused by the RTS/CTS mechanism which is applied in 802.11 MAC to avoid hidden nodes. When a node overhears another transmission and hence refrains to transmit any data of its own, even though such a transmission would not cause a collision due to the limited radio range of the nodes. An exposed node is one that is within the sensing range of the sender but out of the interfering range of the destination. Exposed nodes cause the available bandwidth underutilized. In the 802.11 MAC layer protocol, there is almost no scheme to deal with this problem. This causes a serious problem when it is used in the multi-hop wireless networks [1].

**3.3.5 TCP Aggressive Window Mechanism.** The TCP window mechanism tends to create more signal interference in a wireless ad hoc network environment. It is because the TCP window mechanism drives wireless networks to be crowded with more packets, where a higher spatial density of packets in an area leads to a higher chance of signal interference or collision in a wireless medium. As pointed out in [10], 802.11 MAC cannot perfectly handle signal interference of general multi hop topologies. The push of more packets to go beyond a certain limit by TCP drives excessive link-layer retransmission and eventually leads to more MAC contention loss which further leads to link failure.

Our study shows that MAC link failure can be caused due to different reasons, but IEEE 802.11 MAC treats them in same manner, thus there is a need of mechanism that distinguishes between them and enables MAC to reacts accordingly.

## 4. PROBLEM DEFINITION

MAC 802.11 protocols have been shown to significantly affect MANET performance [11]. Section 3 explains that different interferences are major reason for preventing packets of one node from reaching the other when the two nodes are in each other's transmission range.

As pointed out in Section 2, the IEEE 802.11 MAC protocol reports a link failure if it cannot establish an RTS–CTS handshake with a neighbor node within seven RTS attempts. If a node cannot reach its neighbor node, it drops the packet and triggers a link failure. The MAC protocol fails to establish an RTS-CTS handshake because neighbor node cannot respond to RTS messages within the defined limit due to interference. Thus, with the IEEE 802.11 MAC protocol, *false link failures* may be induced due to interference. A false link failure occurs when the MAC protocol declares that the link to a neighbor node is broken, even though neighbor node is within its transmission range [11].

The on demand routing protocols misinterpret this false link failure in the MAC layer as route failure, triggering the unnecessary route maintenance process, therefore increasing the overhead in the network. When the MAC layer reports a link failure to AODV [12], it simply drops the packets that are to be routed on the failed link. Furthermore, AODV brings down the routes to destinations that include the failed link and sends a route error message to the source of each connection that uses the failed link. Similarly, DSR [13] protocol triggers the route maintenance process and deletes the route entry from its cache and propagate the broken link information to the nodes in its surrounding to bring down their routes to that link.

Routing protocols have to re-compute route to the appropriate destination and if this route is not found in specified time then the TCP sender timesout and invokes its congestion control algorithm. With conventional TCP protocol, when a retransmission timeout happens, the TCP sender retransmits the lost packet and doubles the *Retransmission Time Out* (RTO) period. This procedure is repeated until the lost packet is acknowledged. Such an exponential backoff of the RTO helps TCP react to congestion gracefully. However, when false link failure happens, TCP tends to increase the RTO rapidly even when there is no congestion. Wrongly applied exponential backoff significantly degrades TCP performance [10]. The unnecessary re-routing process interrupts the ongoing TCP traffic flow and greatly degrades the end to end throughput.

The false link failure adversely affects the performance of MANET, thus, it is important to correctly identify such failures. There should be a method which is invoked only when there is a true link failure due to mobility, the proposed algorithm reduces the number of false link failures at MAC layer by adapting the retransmission limit according to network information. We present IEEE 802.11 MAC protocol enhancement that enables to alleviate false link failure. We call our version of the MAC protocol the *Adaptive MAC*. If there is high probability that a node is in transmission range of the sender node, the Adaptive MAC persists in its retransmission attempts and if there is probability that a node has moved away, the Adaptive MAC does not retransmit unnecessary. Our method is a simple cross layer modification to 802.11 MAC protocol because this is the MAC protocol which generates the false link failures and then they are misinterpreted by other higher layers.

## 5. RELATED WORK

There has been significant research on false route failure solutions for MANET [14-17]. In [14], a delayed retransmission (DR) scheme has been proposed by which packets lost in MAC layer are retransmitted in network layer with a delay between two successive transmissions. Further, a delayed adaptive retransmission (DAR) scheme is also proposed in which different retransmission limits are assigned to the packets with different forwarded hops. After transmission failure in DR scheme, the packets are put in tail of sending buffer which makes delay between two successive transmissions. Thus the delay in DR scheme is directly proportional to the packets in sending buffer i.e. longer the queue higher the delay. However, a longer waiting time can be inefficient particularly for TCP flows. Also if the longer time is taken to declare a link as broken then the other packets in the pipeline uses the stale route. The DAR scheme assumes that the topology is stationary and some network parameters are available. However, it does not consider the mobile topology, which is the basic characteristic of MANET.

In [15], another enhancement to the IEEE 802.11 MAC protocol is suggested in which to reduce the false route breakages, additional HELLO messages are sent to the sender whenever the number of RTS received by a receiver exceeds a threshold. This work does not address the effects of the Long Retry Limit, which should intuitively be called into question based on the reason that the link is also declared as failed one if the number of times a node tries to resend the data packet becomes unsuccessful.

A node movement detection scheme is proposed in [16], which allow a node to decide its movement based on the reception status of modified HELLO messages which indicates the degree of its local topology changes. In this scheme, each node periodically broadcasts modified HELLO messages to its neighbors via 1-hop flooding. When a node receives a HELLO message, it updates its neighbor table with message status and makes a prediction on its movement by calculating changes in its neighborhood. Through the mechanism presented in [16], the goal was to decrease overhead due to false route failure, yet the network is flooded with periodic Hello messages whose entries are inserted, updated or deleted into neighbour table according to the reception status of Hello message, which may lead to more overhead than the network began with.

In [17], a route maintenance mechanism between MAC and routing layers, called Congestion Aware Routing (CAR), has been proposed. To reduce false route failure, in CAR, for little or no congestion, nodes report a route as broken at the first transmission failure, but as the monitored congestion increases, nodes slightly dampen route breakage reporting by ignoring a certain number of link failures. It has been shown that CAR can improve throughput in a static topology. The improvement of CAR is achieved based on alleviating the impact of losses by ignoring them in high congestion. This scheme may perform well in static topologies since most packet losses are due to collision. In mobile topologies, however, packets losses are induced not only by collision but also by actual routing failure due to mobility of nodes. Thus, simply ignoring or hiding losses then leads to late respond to routing failure and reduce throughput consequently.

Our approach differs from these efforts in that it integrates cross layer approach in false link failure discovery of MANET. As a result, nodes in our system can exploit available network information to optimize IEEE 802.11 MAC and improve network performance. Our method widens the scope by making both the Short and Long Retry Limits dynamic. Also this method takes into consideration the effects of node own speed to determine the retransmission limits and it deals with the problem of mobility first which the aforementioned papers did not include.

## 6. ADAPTIVE RETRANSMISSION LIMIT (ARL) ALGORITHM

This algorithm adapts the retransmission limits with respect to received signal strength and time of transmitted packet from neighbour node. The objective of this algorithm is to differentiate the false link failures caused by interference from true link failures due to mobility. The key challenge of this algorithm is to distinguish between false and true link failure. In both the cases, a node observes the same result, no CTS within a certain amount of time interval, and it is hard to distinguish them without cross layer interaction.

### 6.1 The flowcharts for ARL Algorithm

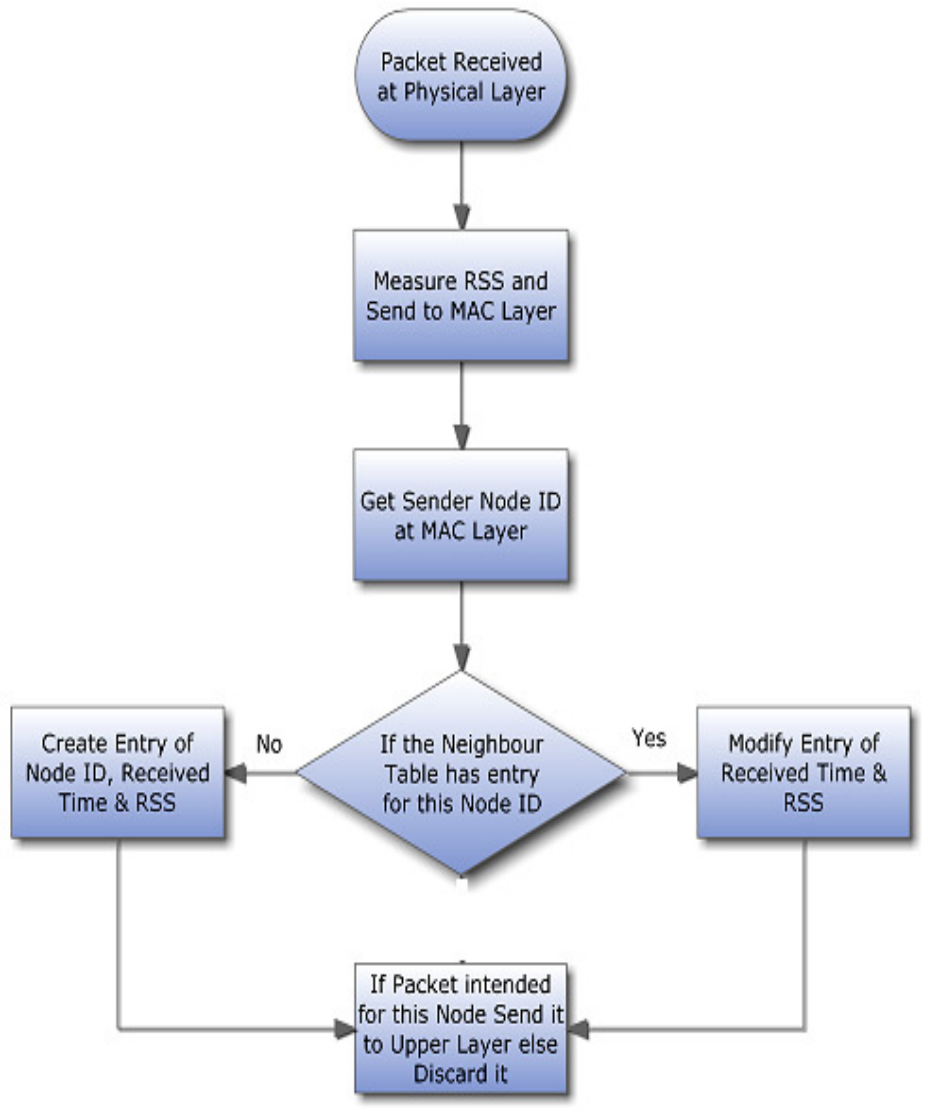

Figure 2. Upon detecting a packet from neighbour node

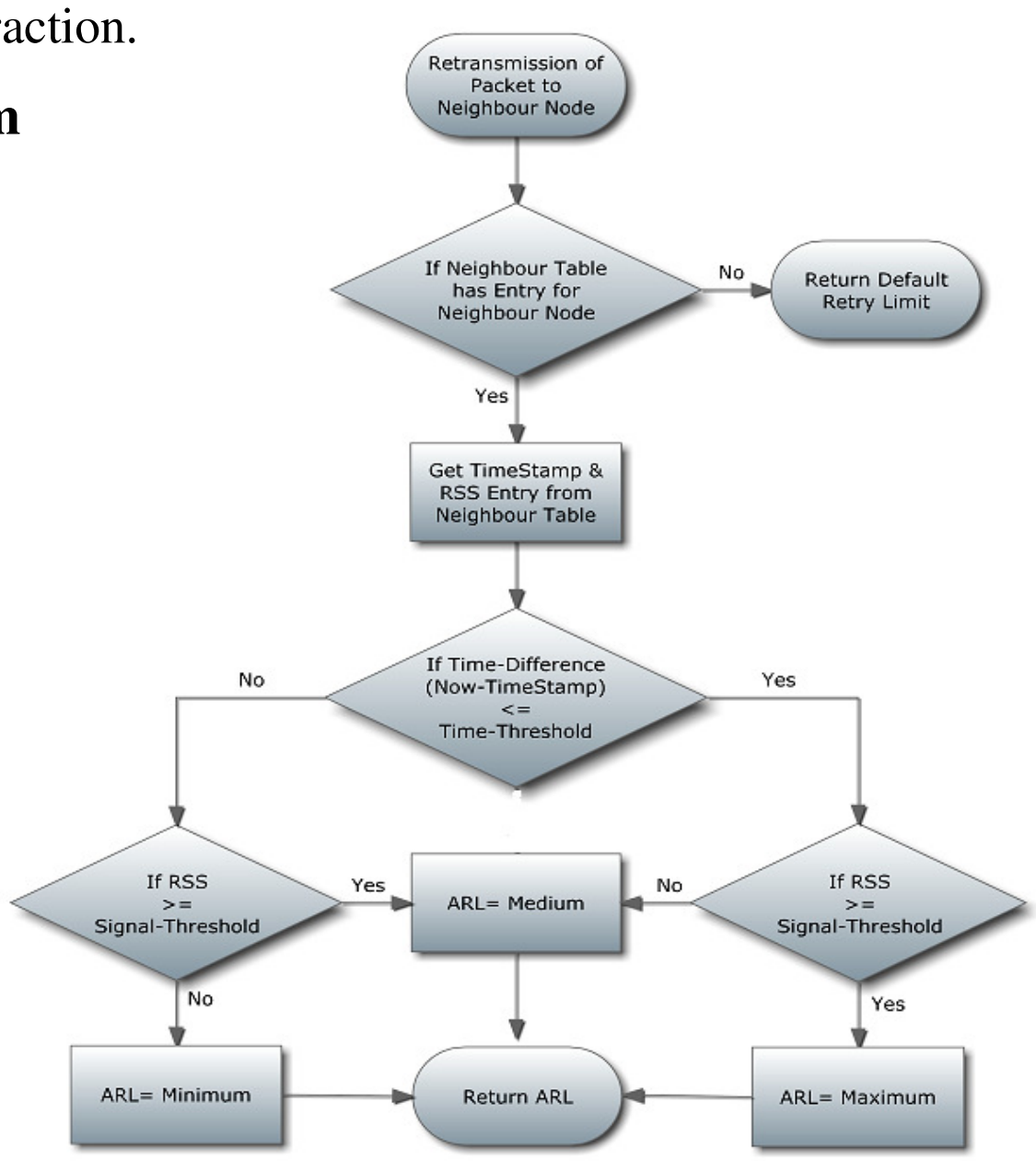

Figure 3. Upon retransmitting a packet to neighbour node

In the proposed algorithm, an adaptive method is employed to distinguish them without additional overhead. It works in two phases. First, each node overhears packets without regard to their destinations, and looks at their source addresses (nodeID) and gets the *Received Signal Strength* (RSS) corresponding to the overheard packet from physical layer, then it stores this information into a table called neighbour table with the receiving time of the packet (timestamp) for later use (Figure 2).

Using the two-ray ground reflection model in ns-2 [18], the RSS at distance d is predicted by

$$P_r(d) = \frac{P_t G_t G_r h_t^2 h_r^2}{d^4 L} \qquad (1)$$

Where, $P_r$ is the received signal strength and $P_t$ is the default transmission strength, $G_t$ and $G_r$ are the antenna gains of the transmitter and the receiver respectively, $h_t$ and $h_r$ are the heights of the antennae, and $L$ is the system loss.

A neighbor table entry (Figure 4) consists of three fields: a nodeID, RSS of the neighbor node packet (estimated using (1)), timestamp at which this packet was received. When a node receives a packet from a same neighbor node, it replaces the old entries of the table, corresponding to that node, with the more recent ones.

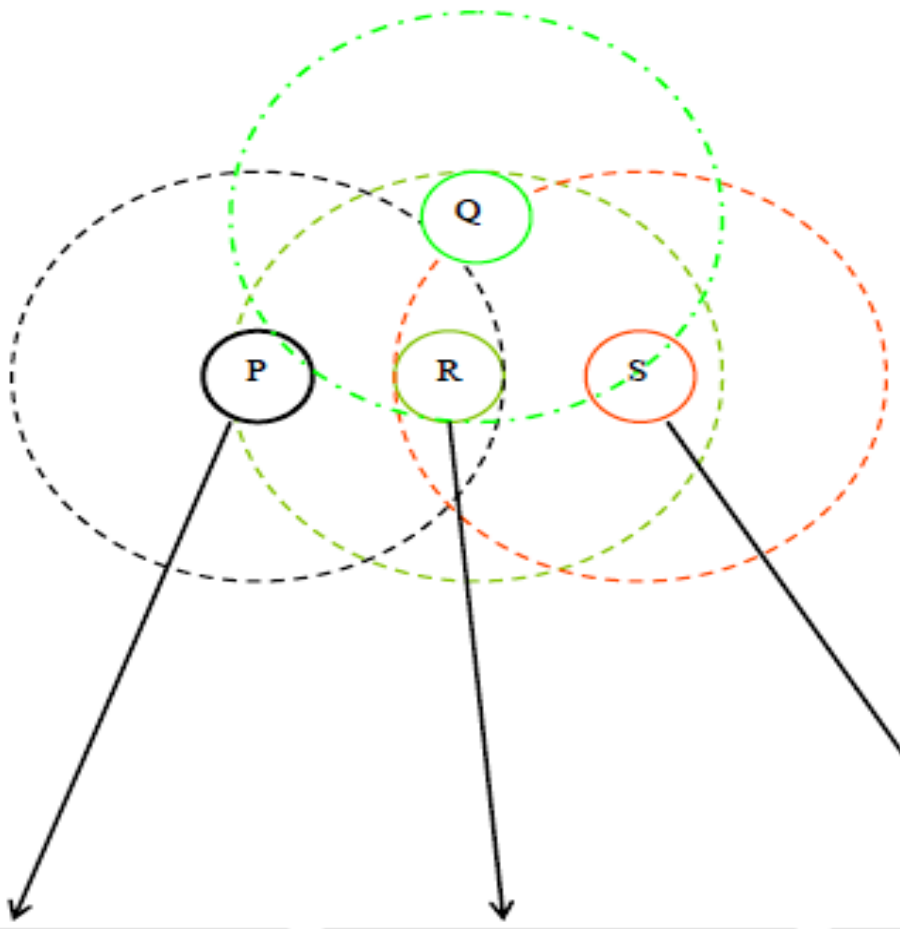

| NodeID | RSS | Timestamp |
|---|---|---|
| R | 9.33E-14 | 13:47:30 |
| Q | 8.74E-14 | 13:47:34 |

Node P Neighbour Table

| NodeID | RSS | Timestamp |
|---|---|---|
| P | 9.75E-14 | 13:47:20 |
| Q | 9.39E-14 | 13:47:31 |
| S | 9.30E-14 | 13:47:35 |

Node R Neighbour Table

| NodeID | RSS | Timestamp |
|---|---|---|
| R | 9.86E-14 | 13:47:25 |
| Q | 8.33E-14 | 13:47:29 |

Node S Neighbour Table

Figure 4. Neighbour Table

The timestamp of overheard packet can be used to detect the broken links due to mobility, because due to mobility the receiver is not reachable and its packets have not been overheard since a long time. The RSS of neighbour nodes is tracked with the purpose of using it later for distinguishing if neighbour nodes are reachable or not i.e. the strong signal indicates that the neighbour node is in transmission range and link will be available for longer time duration and the weak signal indicates that the neighbour node is diminishing and link will soon be broken. Secondly, when a node has to retransmit a RTS or data to the neighbour node, it looks into the neighbour table entry and for the nodes whose packets are overheard recently with high RSS, it applies a large retry limit (Figure 3), since the node is likely to stay nearby. For a node, whose packets have been overheard earlier with low RSS, a small retry limit is applied to avoid unnecessary retransmissions and for the nodes whose packets are overheard recently with low RSS or overheard earlier with high RSS, medium retry limit is applied. Upon transmitting a packet, if a node does not have a table entry for a neighbor node then it uses a default retransmission limit for transmissions to it.

## 6.2 An Algorithm for Adaptive Retransmission Limits

*When a node detects a packet from neighbour node [i]*
nodeID=MAC address
RSS[*i*]= received signal strength {$P_r$ from physical layer}
timestamp[*i*] = the current time {now}

*When a node retransmits a packet to neighbour node [j]*
Now = the current time {now}
timedifference= difference (Now, timestamp[*j*])
if timedifference<=time-threshold
then
if RSS[*j*]>=signal-threshold
ARL[*j*]=maximum[Retry Limit]
if RSS[*j*]<signal-threshold
ARL[*j*]=medium[Retry Limit]
if timedifference>time-threshold
then
if RSS[*j*]>=signal-threshold
ARL[*j*]=medium[Retry Limit]
if RSS[*j*]<signal-threshold
ARL[*j*]=minimum[Retry Limit]

This algorithm returns the *Adaptive Retransmission Limit* (ARL) dynamically in place of static value set by 802.11 MAC. The time-threshold is time required by a node to be out of transmission range if it is moving with a constant speed, which is received from physical layer. Basically, the time-threshold is inversely proportional to the moving speed of the node such that a node moving slowly has a longer time-threshold and a node moving fast has a smaller time-threshold. Thus for the static topology the time-threshold will be very large as the moving speed will be zero. The minimum signal strength (RXThresh) required to receive the packet at physical layer is used to compute the signal-threshold at MAC layer. The signal-threshold of a node is set higher than RXThresh. The maximum, medium and minimum value of retry limit selected for SRL as {16, 12, 4} and for LRL as {8, 6, 2} after the comprehensive evaluation of different values. The Adaptive MAC reports a link failure only if the retransmission limits deduced using the ARL algorithm to establish a handshake or data transmission to neighbour node fails.

## 7. SIMULATION AND ANALYSIS

For the purpose of performance analysis of the proposed algorithm, ns-2 [20] simulation tool has been employed. For our results, we consider two traffic loads, light and heavy. In light traffic load, there are two TCP connections between two different pairs of nodes. The heavy traffic load has eight TCP connections (with data flows in different directions) that cross each other, between eight different pair of nodes. In this simulation, the following performance metrics have been considered for MANET protocols:

- **Link Failures**: The total numbers of link failures reported due to exceeded retry limit.
- **Normalized Routing Load:** The ratio of total routing control packets transmitted and the total data packets received at the destination.
- **Throughput:** The rate of successfully received data packets per second in the network.
- **Average End to End Delay:** The end-to-end-delay is averaged over all received data packets from the sources to the destinations.

The scenario consists of 20 mobile nodes which move in an area of 1000×1000 m according to the random way-point model. The pause time is 5 s and the maximum speed of the mobile nodes is set to 4, 8, 12, 16, 20 and 24 m/s for different simulation runs. The traffic carried by each TCP connection is a continuous file transfer, i.e., a TCP source will send data packets for the entire duration of the simulation. The default transmission range of each node is 250 m and the interference range is 550 m. A TCP packet travels about four hops, on average, to get from a source to the corresponding sink. All TCP sessions last for 300 s.

Table 1 Simulation parameters.

| **Parameter** | **Value** |
|---|---|
| Simulation Time | 300 s |
| No of Mobile Node | 20 |
| Default Transmission Range | 250 m |
| Traffic Type | FTP over TCP |
| Packet Size | 512 Byte |
| TCP window size | 32 |
| No of TCP Connections | 2 in light load, 8 in heavy load |
| Pause Time | 5 s |
| Maximum Speed | 4, 8, 12, 16, 20 and 24 m/s |

## 7.1. Link Failures

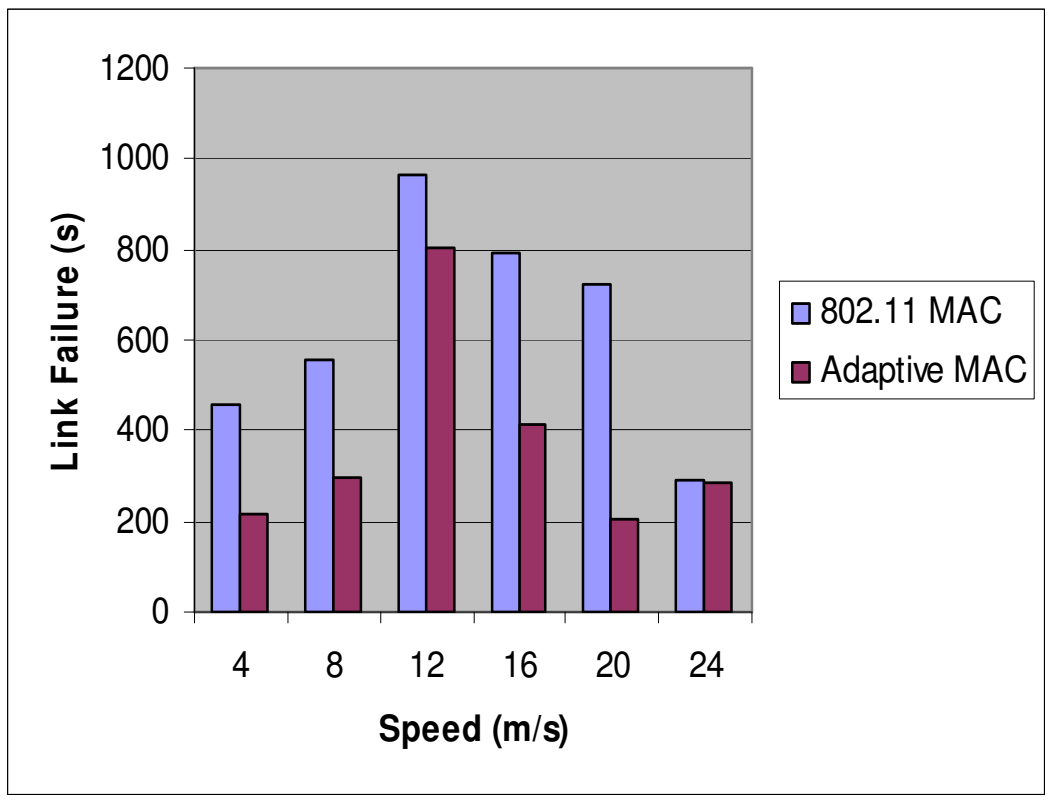

(a) Light load (2 TCP connections)

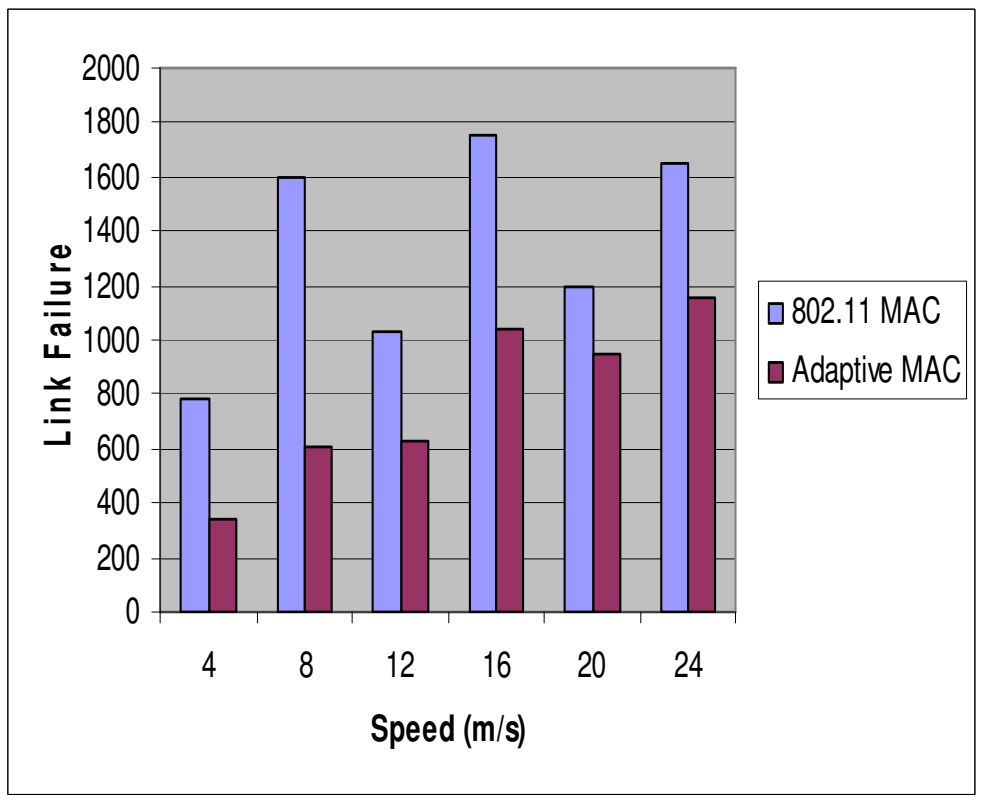

(b) Heavy load (8 TCP connections)

Figure 5. Link Failures

In Figure 5, we can see that the link failures (due to exceeded retry limit) for 802.11 MAC is very high in comparison to Adaptive MAC, the reason behind this difference is that the 802.11 MAC does not differentiate between true and false link failures as compared to the Adaptive MAC which persists in its retransmission attempts and declares the link failure only when the node is out of transmission range.

## 7.2. Normalized Routing Load

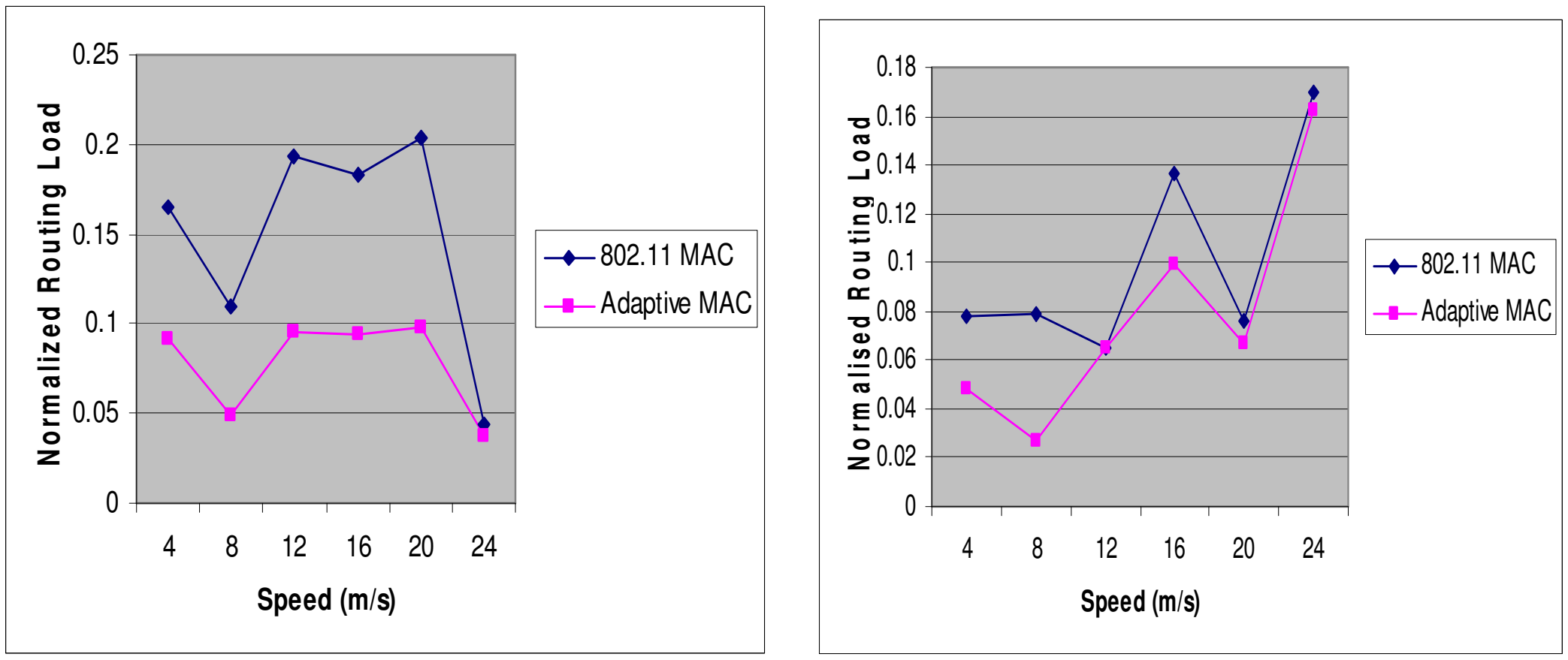

(a) Light load (2 TCP connections) (b) Heavy load (8 TCP connections)

Figure 6. Normalized Routing Load

Figure 6 shows how routing overhead is severely reduced, since there is less routing activity at routing layer using Adaptive MAC. We see from Figure 5 that link failures are decreased for all speeds and both loads for Adaptive MAC which causes the less routing activity and thus reduced Normalized Routing Load. Usually, in random waypoint scenarios, there are high link failures because the link failures over multiple hops are due to node mobility (true link failure) and due to interference (false link failure). The Adaptive MAC handles the failure appropriately and causes the route rediscovery only when it required, reducing the routing load.

## 7.3. Throughput

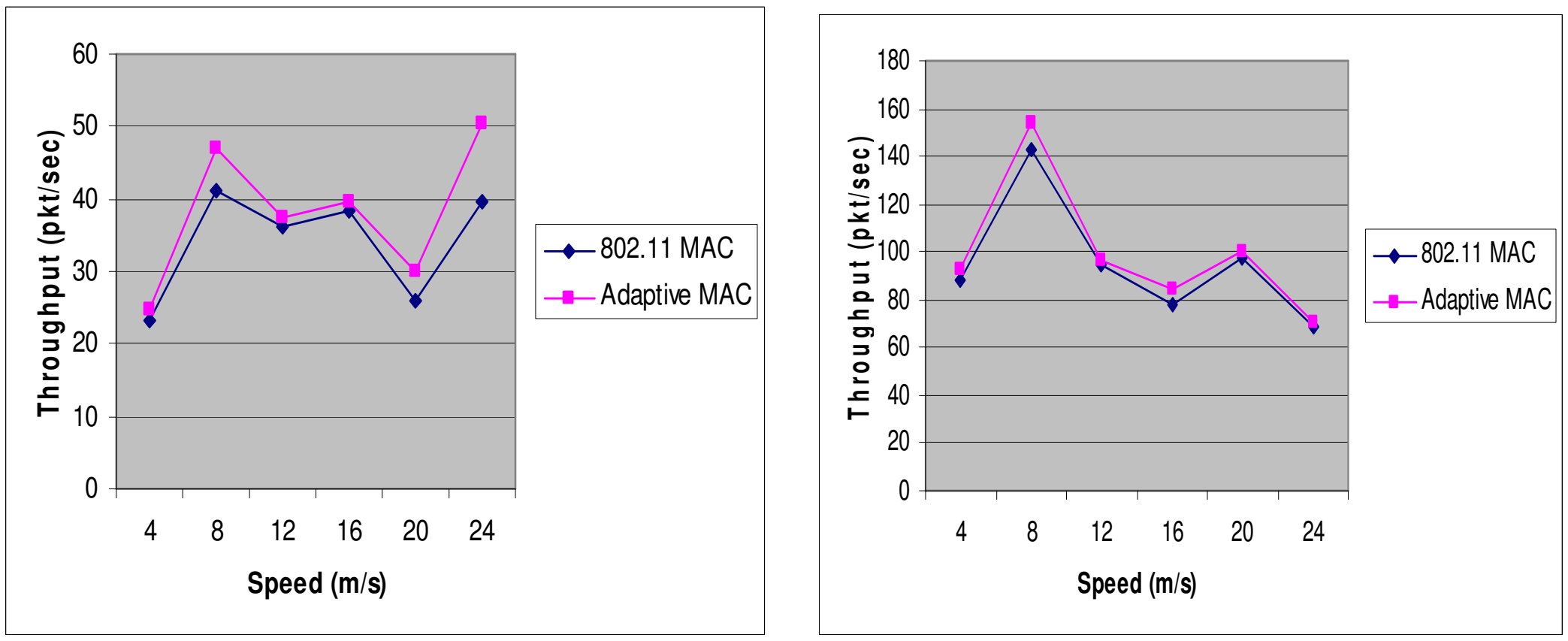

(a) Light load (2 TCP connections) (b) Heavy load (8 TCP connections)

Figure 7. Throughput

As discussed in section 7.1 & 7.2, 802.11 MAC reports more link failures as compared to Adaptive MAC (Figure 5). Due to these false link failures, routing protocol causes a high number of route errors leading to more routing overhead (Figure 6) thus decreased level of bandwidth available for data packets. Further, the routing protocol drops the data packet due to false reporting. Both of these condition lead to decreased throughput level in 802.11 MAC. Therefore, throughput increases by using Adaptive MAC as also confirmed by the results of simulation shown in Figure 7.

### 7.4. Average End to End Delay

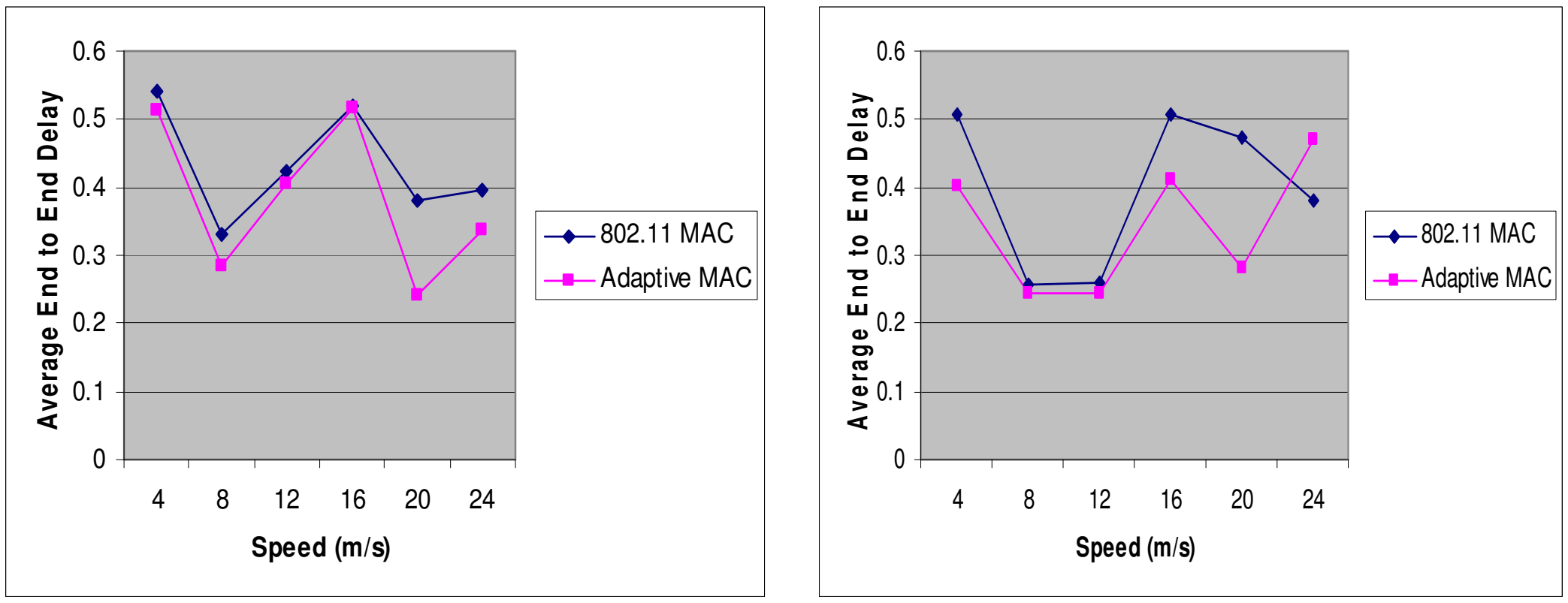

(a) Light load (2 TCP connections) (b) Heavy load (8 TCP connections)

Figure 8. Average End to End Delay

As is evident from the results shown in Figure 8, in the case of light load condition, average end-to-end delay is less in Adaptive MAC as compared to 802.11 MAC, while in the case of heavy load condition, the delay increases with increased speed of mobile nodes and thus for delay sensitive application the Adaptive MAC performs well in case of light load scenario.

## 8. Conclusion and Future Work

In this paper our objective is to reduce the false link failure in mobile ad hoc networks and thereby improve their performance. Towards this, we propose an adaptive retransmission limit algorithm that helps in minimizing false link failures. This algorithm is based on cross layer interaction which gathers the physical layer statistics and uses, in MAC layer, to determine if a node is in transmission range or not.

With our extension, the Adaptive MAC dynamically selects the number of retransmission attempts in order to establish a link. Based on signal strength measurements and time of overhearing packets, the maximum attempts are performed for strong signal nodes to avoid the false link failures. Further, our algorithm identifies weak signal nodes, and selects the minimum number of retransmission attempts which, in turn, helps in switching to the new route even before the link failure occurs in 802.11 MAC.

The simulation results show that Adaptive MAC can considerably improve the performance of MANET. It reduces the link failures which cause less route failure and lower routing overhead in the network. As the routing overhead is decreasing, the nodes are able to transmit more data packets, consequently, the number of TCP retransmission timeout is reduced and the TCP source sends more packets, therefore, a higher throughput is obtained.

Future work involves exploiting cross layer interaction in reactive routing protocols such as DSR or AODV. For example, DSR can use physical layer statistics to determine link quality and react accordingly.

## References

[1] Imrich Chlamtac, Marco Conti and Jennifer J.-N. Liu. “Mobile ad hoc networking: imperatives and challenges,” Ad Hoc Networks, vol. 1, pp. 13–64, 2003.

[2] V. Srivastava and M. Motani. “Cross-layer design: a survey and the road ahead,” *IEEE Communication Magazine,* vol. 43, no. 12, pp. 1112–119, 2005.

[3] Fuad Alnajjar, Yahao Chen, "SNR/RP aware routing algorithm: cross-layer design for manets," International Journal of Wireless & Mobile Networks (IJWMN), Vol 1, No 2, November 2009.

[4] X. Yin. "Improving tcp performance over mobile ad hoc networks by exploiting cross-layer information awareness," in *Proc. The 10-th annual international conference on Mobile computing and networking,* Philadelphia, USA, pp. 231–244, 2004.

[5] Kitae Nahm, Ahmed Helmy, and C.-C. Jay Kuo, "Cross-layer Interaction of TCP and Ad Hoc Routing Protocols in Multihop 802.11 Networks," *IEEE Transactions on Mobile Computing (TMC)*, vol. 7, no. 4, pp. 458 - 469, 2008.

[6] IEEE Computer Society, "IEEE Standard for Wireless LAN Medium Access Control (MAC) and Physical Layer (PHY) Specifications", *International Standard ISO/IEC* 8802-11: 1999(E), ANSI/IEEE Std 802.11, 1999 Edition.

[7] J. Li, C. Blake, D. S. J. De Couto, H. Lee, and R. Morris, "Capacity of ad hoc wireless networks," in *Proc. ACM MobiCom*, Rome, Italy, 2001.

[8] K. Xu, M. Gerla, and S. Bae, "How Effective is the IEEE 802.11 RTS/CTS Handshake in Ad Hoc Networks?" in *Proc. GLOBECOM*, 2002.

[9] V.Shah, S.V.Krishnamurthy and N.Poojary, "Improving MAC Layer Performance in Ad Hoc Netowrks of Nodes with Heterogeneous Transmit Power Capabilities," in *Proc. ICC, 2004.*

[10] Z. Fu, P. Zerfos, H. Luo, S. Lu, L. Zhang, and M. Gerla, "The impact of multihop wireless channel on TCP throughput and loss," in *Proc. IEEE INFOCOM*, San Francisco, CA, 2003.

[11] Shugong Xu, Tarek Saadawi, "Revealing the problems with 802.11 medium access control protocol in multi-hop wireless ad hoc networks," *Computer Networks*, vol. 38, pp. 531-548, 2002.

[12] C. E. Perkins, E. M. Belding-Royer, S. R. Das, "Ad Hoc On-Demand Distance Vector (AODV) Routing", *IETF MANET Working Group INTERNET DRAFT*, 19 January 2002.

[13] D. B. Johnson, D. A. Maltz, Y.-C. Hu, "The Dynamic Source Routing Protocol for Mobile Ad Hoc Networks (DSR)", *IETF MANET Working Group INTERNETDRAFT*, 19 July 2004.

[14] Qing Chen Zhisheng Niu, "A delayed adaptive retransmission scheme for false route failure in MANET"; Proceedings of the 2004 Joint Conference of the 10th Asia-Pacific Conference on Communications, 2004, Vol. 2, pp. 858 – 862.

[15] Xia Li Kee-Chaing Chua Peng-Yong Kong; "The Study of False Route Breakage in IEEE 802.11 based Ad Hoc Networks"; Proc. of the IEEE International Conference on Mobile Adhoc and Sensor Systems (MASS), 2006, pp.: 493 – 496.

[16] Hyun Yu Sanghyun Ann "Node Movement Detection to Overcome False Route Failures in Mobile Ad Hoc Networks", in International Conference on Information Science and Security, Seoul 2008.

[17] U. Ashraf, S. Abdellatif and G. Juanole "Efficient Route Maintenance in Wireless Mesh Networks", in 3rd International Symposium on Wireless Pervasive Computing (IEEE ISWPC 2008) Santorini, Greece May 2008.

[18] Information Sciences Institute (ISI).*The Network Simulator – ns-2.* http://www.isi.edu/nsnam/ns/